\documentclass{eptcs}
\providecommand{\event}{ADG 2021} 
\usepackage{graphicx}
\usepackage{hyperref}

\usepackage[utf8]{inputenc}
\usepackage[T1,OT1]{fontenc}
\usepackage[inline]{enumitem}
\setlist{nosep}
\usepackage{tikz}
\usepackage{varwidth}

\newtheorem{example}{Example}

\newenvironment{figures}[4][H]
{ \refstepcounter{figure}
\begin{center}
\setlength{\unitlength}{#1 mm} {\small
\resizebox{#1\height}{!}{\input{#2}}
} \setlength{\unitlength}{1.0 mm}
\end{center}
\label{#4} \centerline{{\bf Figure \thefigure:} #3}
\par
\vspace{0.5em} \noindent }{}

\newenvironment{figures*}[2][H]
{
\begin{center}
\setlength{\unitlength}{#1 mm} {\small
\resizebox{#1\height}{!}{\input{#2}}
} \setlength{\unitlength}{1.0 mm}
\end{center}
\par
\vspace{0.5em} \noindent }{}

\def\orcidID#1{\unskip$^{\mbox{\href{https://orcid.org/#1}{\scriptsize{[#1]}} }}$}
\begin{document}

\title{On Automating Triangle Constructions in Absolute and Hyperbolic
  Geometry\thanks{This research is supported by the Serbian Ministry
    of Education, Science and Technological Development through the
    University of Belgrade, Faculty of Mathematics.}}

\author{Vesna Marinković\orcidID{0000-0003-0526-899X} 
\institute{Faculty of Mathematics \\ University of Belgrade, Serbia}
\email{vesnap@matf.bg.ac.rs}
\and
Tijana Šukilović\orcidID{0000-0001-6371-3081}
\institute{Faculty of Mathematics \\ University of Belgrade, Serbia}
\email{tijana@matf.bg.ac.rs}
\and
Filip Marić\orcidID{0000-0001-7219-6960}
\institute{Faculty of Mathematics \\ University of Belgrade, Serbia}
\email{filip@matf.bg.ac.rs}
}

\def\titlerunning{On Automating Triangle Constructions in Absolute and Hyperbolic Geometry}
\def\authorrunning{Marinković, Šukilović \& Marić}

\maketitle

\begin{abstract}
  We describe first steps towards a system for automated triangle
  constructions in absolute and hyperbolic geometry. We discuss key
  differences between constructions in Euclidean, absolute and
  hyperbolic geometry, compile a list of primitive constructions and
  lemmas used for constructions in absolute and hyperbolic geometry,
  build an automated system for solving construction problems and test
  it on a corpus of triangle-construction problems. We also provide an
  online compendium containing construction descriptions and
  illustrations.  
\end{abstract}


\section{Introduction}

Ruler\footnote{Usually only a straightedge (ruler with no marks) is
  allowed.} and compass triangle constructions
(abbr.~RC-constructions) have been intensively studied in mathematics
and mathematical education since ancient Greeks, for now more than two
and a half thousand years. Triangles are also a very nice polygon to
develop and test artificial intelligence and automated reasoning
tools, and systems that automatically solve RC-construction problems
are actively being developed (e.g., ArgoTriCS~\cite{argotrics}).

Construction problems may require constructing a triangle given some
of its significant points. William Wernick~\cite{wernick-corpus} has
given an exhaustive catalogue of all such problems formulated over the
following significant points: three vertices $A$, $B$, $C$,
circumcenter $O$, three side midpoints $M_a$, $M_b$, $M_c$, centroid
$G$, three feet of altitudes $H_a$, $H_b$, $H_c$, orthocenter $H$,
three feet of internal angle bisectors $T_a$, $T_b$, $T_c$, and
incenter $I$ (see Figure~\ref{fig:wernick_points})\footnote{This
  notation will be used throughout the paper}. All problems require
constructing a triangle given its three distinct significant points
(from those listed). There are 139 such problems that are
significantly different. Some problems are redundant (e.g., the triple
$A$, $B$, and $M_c$ is redundant since the midpoint $M_c$ can be
constructed when $A$ and $B$ are given). Some problems are locus
dependent (e.g, the triple $A$, $B$, and $O$ is solvable only if $O$
belongs to the perpendicular bisector of the segment $AB$). When
redundant and locus dependent problems are excluded, there remain 114
problems, and Wernick solved 65 of these, leaving the others with an
unknown status. In the meantime status of all problems from the
Wernick's list has been determined, by either solving them, or proving
that required triangles are not constructible by using only ruler and
compass: there are 74 solvable problems, 39 unsolvable, 3 redundant
and 23 locus dependent problems~\cite{wernick-final-update}.

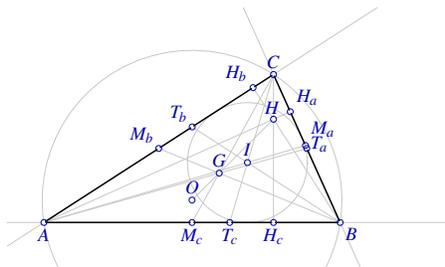
\begin{figures}[0.85]{points_w.tkz}{Significant points from Wernick's
    corpus.}{fig:wernick_points}\end{figures}

RC-constructions are usually done in the framework of Euclidean
geometry. Triangle constructions in other geometries (e.g.,
hyperbolic) have been studied, but to much lesser extent. In this
paper we investigate automated triangle constructions in absolute and
hyperbolic geometry (focusing on the Poincar\'e disc model). We
describe an automated solver for those problems, based on our
Euclidean triangle construction solver ArgoTriCS. We describe all
changes that have been necessary to adapt ArgoTriCS to fully
automatically find triangle constructions in absolute and hyperbolic
geometry. We apply the modified system to Wernick's corpus and
summarize the current results.

\subsection{Related work}

\textit{Constructions in hyperbolic geometry.}  The theory of
hyperbolic constructions was almost entirely developed by
Russian-writing mathematicians (e.g., Mor\-dou\-khay-Boltovskoy~\cite{mb-lobacevski}).
Pambuccian gives constructive axiomatizations
of several geometries~\cite{pambuccian-survey}, including Euclidean,
absolute and hyperbolic geometry. Unlike traditional geometry
axiomatizations (e.g., Hilbert's and Tarski's) that use only relation
and no function (operation) symbols, Pambuccian does quite the
opposite and gives axiomatic systems that use only function and no
relation symbols. There are several toolboxes in dynamic geometry
systems (GeoGebra, Geometer's Sketchpad, Cinderella, etc.) that
facilitate manual step-by-step constructions in Poincar\'e disc model.

\textit{Automated triangle constructions.}  Despite long tradition of
RC-constructions, there are quite few systems that automate
them. Schreck developed Prog\'e~\cite{proge} -- a general framework
implemented in PROLOG where different kind of geometric objects (not
only triangles) can automatically be constructed. Gao and Chou applied
algebraic approach and used Wu's method~\cite{wu} or Gr\"obner
bases~\cite{groebner} to find locations of unknown objects from the
locations of known objects and determine
RC-constructibility~\cite{gaochou}. Schreck~\cite{schreck-mechanization}
also focused on the algebraic approach and compared Gao and Chou's
method with the Lesbegue's method using them to show
non-constructibility, but also to extract some RC-constructions from
algebraic methods. Gulwani et al.~\cite{gulwani} formulated a logic
and a programming language of geometry constructions and then applied
methods for program synthesis to obtain RC-constructions. Marinković
and Janičić~\cite{Marinkovic12,argotrics} focused on triangle
construction given its significant points. Their system ArgoTriCS
is capable of solving 66 out of 74 solvable problems from Wernick's
list. It can also detect all redundant and locus dependent problems
from this list.

As far as we know, there has been no previous research on automated
triangle constructions or automated theorem proving in non-Euclidean
geometries. Algebraic approach that is very successful for Euclidean
geometry is hard to apply in the hyperbolic setting since distance
constraints are not describable using only polynomials (distances
involve logarithms of cross-ratio).

\section{ArgoTriCS}

ArgoTriCS is a PROLOG system that can solve geometry construction
problems automatically, given some background geometrical
knowledge~\cite{argotrics}. It generates informal description of found
construction in natural language (in English) and formal description
in GCLC language~\cite{gclc-jar}, accompanied by a corresponding
illustration. It also generates non-degeneracy conditions which
guarantee that the obtained solution exists and has support for
proving that generated constructions are correct by using
OpenGeoProver~\cite{GATPFormalization} and provers available within
GCLC tool.

The knowledge base that the system requires was identified through a
careful analysis of constructions available in the literature. We came
up to a core geometrical knowledge needed for solving triangle
construction problems from Wernick's
corpus. 
This knowledge was split into three sets: set of definitions, set of
lemmas, and set of primitive constructions. We distinguish between
instantiated and general definitions and lemmas. For instance, in a
given triangle there is only one orthocenter, one centroid, and one
circumcenter. Therefore the lemma of Euclidean geometry stating that
these three points are collinear and that the centroid is between
circumcenter and orthocenter and is twice as far from the orthocenter
than from the circumcenter is an instantiated lemma. On the other
hand, the lemma stating that the center of a circle belongs to the
bisector of an arbitrary chord of that circle is a general lemma, and
therefore it is applicable to any chord of any circle.  However,
before the search for a construction starts, general definitions are
instantiated with all relevant objects and added to the knowledge base
as derived definitions, and, similarly, general lemmas are
instantiated with objects that satisfy their preconditions and added
to the knowledge base as derived lemmas.  Primitive constructions used
in the system are always non-instantiated.

The search procedure starts from the given points and tries to
construct all three vertices of the triangle. Primitive constructions
are used to construct new objects from the current ones: if some
primitive construction can be instantiated so that all objects from
its preconditions are already constructed, but its resulting object is
not yet constructed, then the object is added to the set of
constructed objects. Let us consider the primitive construction saying
that for two given points one can construct the line passing through
them; if, for instance, the circumcenter $O$ of the triangle $ABC$ and
the midpoint $M_a$ of the side $BC$ are already constructed, but not
the perpendicular bisector $m_a$ of the side $BC$, we can use this
primitive construction, construct the line $m_a$ and add it to the set
of constructed objects. After successful application of one primitive
construction, the search for the next applicable primitive
construction starts from the beginning. If all three vertices of
triangle are constructed, the search succesfully finishes. If no
primitive construction is applicable and at least one of the vertices
is not yet constructed, the problem is declared unsolvable using given
knowledge base (however, it does not mean that it is not constructible
given ruler and compass). Some additional techniques are employed in
order to make the search process more efficient. For instance, only
relevant objects are constructed -- a midpoint of a segment determined
by two known points is constructed only if it appears in a definition
or a lemma involving some object, not yet constructed.  Therefore,
definitions and lemmas guide the construction process. After the
construction is automatically generated, it is simplified and all
irrelevant steps that are not used for the construction of triangle
vertices are eliminated.

\begin{figures}[0.5]{construction0_0031.tkz}{\begin{varwidth}[t]{\linewidth}
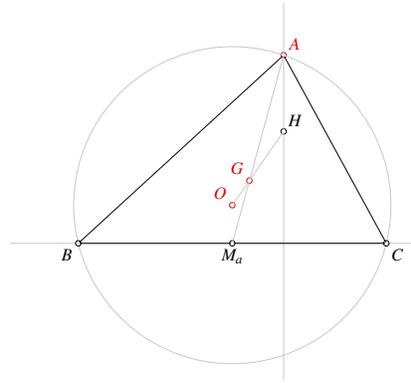
       Construction of triangle $ABC$ in Euclidean geometry\\
       given the vertex $A$, circumcenter $O$, and centroid $G$.
     \end{varwidth}}{fig:construction_0031}
\end{figures}

\begin{example}\label{ex:euclid}
  Let's take a look at one construction automatically found by
  ArgoTriCS. The problem is to construct triangle $ABC$ given vertex
  $A$, circumcenter $O$ and its centroid $G$. The sequence of
  construction steps is the following (see Figure
  \ref{fig:construction_0031}):

  \begin{enumerate}[wide, topsep=0pt, itemsep=1pt, labelwidth=!,
    labelindent=0pt, leftmargin=3\parindent, label=\textbf{Step
      \arabic*.}]
  \item Construct the point $M_a$ for which holds
    $\overrightarrow{AM_a}/\overrightarrow{AG}=3/2$;
  \item Construct the point $H$ for which holds
    $\overrightarrow{OH}/\overrightarrow{OG}=3$;
  \item Construct the line $h_{a}$ through the points $A$ and $H$;
  \item Construct the circle $\kappa$ centered at point $O$ passing
    through point $A$;
  \item Construct the line $a$ perpendicular to the line $h_{a}$
    passing through point~$M_{a}$;
  \item Construct the intersection points $B$ and $C$ of the circle
    $\kappa$ and the line~$a$.
  \end{enumerate}

  The facts used for generating this solution are that centroid $G$
  divides the median $AM_a$ in the ratio $2:1$ and that centroid $G$
  is between circumcenter $O$ and orthocenter $H$ and twice as far
  from the orthocenter than from the circumcenter.
\end{example}

Solving time of different construction problems differ a lot: for most
problems it is a couple of milliseconds, while for some (e.g. locus
dependent problems) it can last more than an hour. Most of the
generated constructions are the same as the ones found in literature.

\subsection{Primitive constructions for Euclidean geometry}\label{subsec:primconstE}

Here we list a few primitive constructions that ArgoTriCS uses for
generating constructions in the framework of Euclidean geometry. Note
that high-level constructions (such as dropping a perpendicular or
constructing a parallel) can be expressed in terms of basic, low-level
RC-constructions.

\begin{enumerate}[wide, topsep=0pt, itemsep=1pt, labelwidth=!, labelindent=0pt]
\item Given points $X$ and $Y$ one can construct a line $XY$;
\item Given two distinct points $X$ and $Y$ it is possible to
  construct a circle
  centered at point $X$ which passes through the point $Y$;
\item Given two lines, it is possible to construct their intersection
  point;

\item Given a point $X$ and a line $p$ one can construct a line $q$
  which passes through the point $X$ and which is perpendicular to the
  line $p$;
\item Given points $X$ and $Y$ one can construct the bisector of the
  segment $\overline{XY}$;


\item\label{it:w10E} Given a point $X$ and a line $p$ one can
  construct a line which passes through the point $X$ and which is
  parallel to the line $p$;


\item\label{it:w01E} Given points $X$, $Z$, and $W$, and a rational
  number $r$ (given by its numerator and denominator) one can
  construct a point $Y$ for which holds:
  $\overrightarrow{XY}/\overrightarrow{ZW}=r$.
\end{enumerate}

\subsection{Definitions and Lemmas for Euclidean geometry}

Let's take a look at the couple of definitions used by ArgoTriCS.
\begin{enumerate}[wide, topsep=0pt, itemsep=1pt, labelwidth=!, labelindent=0pt]
\item Circumcenter $O$ is the intersection point of the perpendicular bisectors of the
  segments $BC$ and $AC$;

\item Orthocenter $H$ is the intersection point of the altitudes $h_a$ and
  $h_b$;


\item Feet of altitudes $H_a$, $H_b$, and $H_c$ are intersection
  points of altitudes with opposite sides of triangle;

\item Circumcircle is the circle centered at circumcenter passing through point $C$.

\end{enumerate}

Here we list a few instantiated and one general lemma used for carrying
out constructions in Euclidean geometry and used by ArgoTriCS.

\begin{enumerate}[wide, topsep=0pt, itemsep=1pt, labelwidth=!, labelindent=0pt]
\item\label{le:1E} Points $C$, $H_b$, and $H_c$ belong to the circle
  centered at point $M_a$ passing through point $B$; points $A$,
  $H_a$, and $H_c$ belong to the circle centered at point $M_b$
  passing through point $C$; points $B$, $H_a$, and $H_b$ belong to
  the circle centered at point $M_c$ passing through point $A$;
\item\label{le:2E} $\overrightarrow{AG}/\overrightarrow{AM_a}= 2/3$,
  $\overrightarrow{BG}/\overrightarrow{BM_b}= 2/3$,
  $\overrightarrow{CG}/\overrightarrow{CM_c}= 2/3$;


\item\label{le:midlineE} Lines $M_aM_b$ and $AB$ are parallel; lines
  $M_aM_c$ and $AC$ are parallel; lines $M_bM_c$ and $BC$ are
  parallel;


\item\label{le:6E} $\angle HAI=\angle IAO$, $\angle HBI=\angle IBO$,
  $\angle HCI=\angle ICO$;
\item\label{le:7E} Center of an arbitrary circle belongs to the
  bisector of an arbitrary chord of that circle.
\end{enumerate}

\section{Triangle constructions in non-Euclidean geometries}

In the rest of the text we shall assume the basic knowledge about
hyperbolic geometry and its models (e.g., Poincar\'e disk model). For
more details on hyperbolic geometry and its relations to Euclidean
geometry we refer the readers to classic
literature~\cite{coxeter,milnor}. Also, for the avid readers, we
strongly recommend reading the original works of Lobachevsky and
Bolyai.

First, let us emphasize the relation between absolute geometry and the
Euclidean or hyperbolic setting. It is well known that the absolute
geometry is based on four groups of axioms: incidence, order,
congruence, and continuity. By adding the appropriate axiom of
parallelism, we get either Euclidean or hyperbolic geometry.

Recall that the question of parallelism is basically the following: In
Euclidean geometry there is a unique line parallel to a given line $a$
through a point $A$ not on the line, while in hyperbolic geometry
there are infinitely many parallels to $a$ through a point
$A$. However, there are exactly two parallel lines that contains the
limiting parallel rays (in Poincar\'{e} disc model those are h-lines
meeting only at the infinity, i.e. on the absolute) and those two
lines we call \emph{parallel} to a line $a$. All the other lines not
intersecting line $a$ are called \emph{hyperparallel}. For example, a
line through point $A$ that is parallel/hyperparallel to line $BH_b$
is the line $AH_a$ from the middle/right picture of
Figure~\ref{fig:pencil}, respectively.

Many statements related to the geometry of a triangle can be proved without
a notion of parallelism. Let us mention some of the main ones:
\begin{enumerate}[wide, topsep=0pt, itemsep=1pt, labelwidth=!, labelindent=0pt]
\item The sum of internal angles of a triangle is less or equal to
  $\pi$.
\item The three medians of a triangle intersect in one point (the
  centroid $G$).
\item The three internal angle bisectors of a triangle intersect in
  one point (the incenter $I$).
\item The perpendicular bisectors of triangle sides belong to the same
  pencil of lines (see~\cite{coxeter}). The same holds for the altitudes of a triangle.
\end{enumerate}

In the Euclidean case, the sum of the internal angles of each triangle
is exactly $\pi$, the perpendicular bisectors of triangle sides are
concurrent lines that intersect in the circumcenter $O$ and altitudes
are also concurrent and intersect in the orthocenter $H$. In the
hyperbolic case, the sum of the internal angles of each triangle is
always less than $\pi$. Also, unlike the Euclidean case, similar
triangles are always congruent. As illustrated in
Figure~\ref{fig:pencil}, in the hyperbolic case the altitudes of the
triangle may not intersect. The same holds for the perpendicular
bisectors. Namely, altitudes/perpendicular bisectors of triangle
belong to the same pencil of lines. In the Euclidean case, this pencil
is always elliptic, i.e. the lines are concurrent, but in the
hyperbolic case, it can also be parabolic (lines are parallel) or
hyperbolic (lines are hyperparallel, i.e. they poses a common
perpendicular).

\begin{minipage}{.3\textwidth}
\begin{figures*}[0.45]{ortho_ell.tkz}\end{figures*}
\end{minipage}
\begin{minipage}{.3\textwidth}
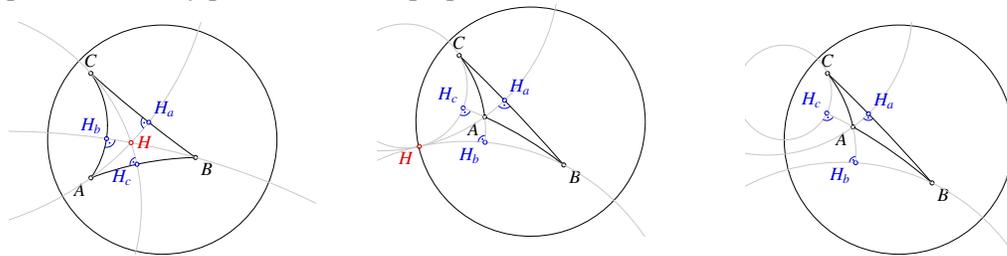
\begin{figures}[0.45]{ortho_par.tkz}{Elliptic, parabolic, and
      hyperbolic pencil of altitudes.}{fig:pencil}\end{figures}
\end{minipage}
\begin{minipage}{.3\textwidth}
  \begin{figures*}[0.45]{ortho_hyp.tkz}\end{figures*}
\end{minipage}

Therefore, one can expect many similarities, but also many differences
between constructions in Euclidean and hyperbolic geometry.

\textit{Hyperbolic instruments.} Ruler and compass can be used to draw
points, straight lines, and circles, and these are the most important
loci of points in Euclidean geometry. For example, the set of points
on one side of a line, equidistant from that line forms another line
(parallel to the original). All vertices of angles of given size
subtending a line segment lie on a circular arc. In hyperbolic
geometry there are other types of curves that can be of interest. For
example, the equidistant curve is not a line, but a \emph{hypercycle}
(Figure \ref{fig:hyper}). Another useful curve type is
\emph{horocycle} (Figure \ref{fig:horo}) whose normal lines all
converge asymptotically in the same direction (in the Euclidean case
horocycles are lines, and all normal lines of a horocycle are parallel
and thus converge in the same direction).

\begin{minipage}{.45\textwidth}
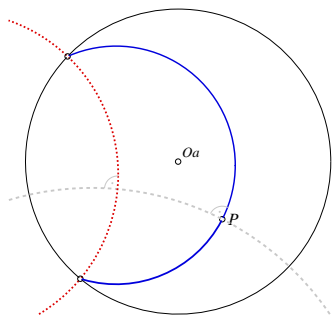
  \begin{figures}[0.5]{hypercycle.tkz}{\begin{varwidth}[t]{\linewidth}Hypercycle (solid blue) through\\ point $P$ with its axis (dotted red)\\ and normal (dashed)\end{varwidth}}{fig:hyper}\end{figures}
\end{minipage}
\begin{minipage}{.45\textwidth}
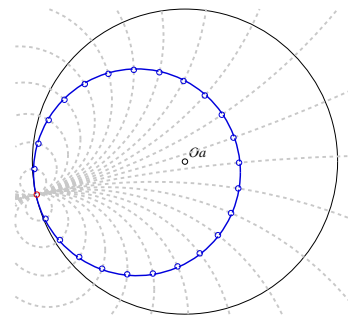
  \begin{figures}[0.5]{horocycle.tkz}{\begin{varwidth}[t]{\linewidth}Horocycle (solid blue) and \\ its perpendicular\\ lines (dashed)\end{varwidth}}{fig:horo}\end{figures}
\end{minipage}

Therefore, it is very important to make precise what instruments can
be used in a construction and what curves can be constructed. Various
hyperbolic instruments have been
proposed~\cite{hyperbolic-instruments}. The \emph{hyperbolic ruler}
draws a h-line given its two different points, the \emph{hyperbolic
  compass} draws a h-circle given its center and a point, the
\emph{hypercompass} draws a hypercycle given its central line and
radius, and the \emph{horocompass} draws a horocycle through a given
point, given its diameter through the point with its direction. An
important (and a bit surprising) result in the theory of hyperbolic
constructions shows that in conjunction with a hyperbolic ruler, all
three compasses are equivalent (everything that can be constructed by
a hyperbolic ruler and one of these three compasses can also be
constructed by a hyperbolic ruler and any other one of these three
compasses)~\cite{mb-lobacevski,coxeter}. Using a combination of a
hyperbolic ruler and any of the three compasses one can also draw a
h-line through a given point parallel to a given h-ray and a h-line
parallel to one and perpendicular to the other given
h-line~\cite{handest,hyperbolic-instruments}.

Additional confusion can come from the fact that in some models of
hyperbolic geometry circles represent many different hyperbolic
curves. For example, in the Poincar\'e disc model h-lines are
represented by Euclidean circular arcs that are orthogonal to the unit
circle (absolute), h-circles are represented by Euclidean circles
fully contained within the absolute (however, the h-center does not
need to match the Euclidean center), hypercycles are represented by
Euclidean circular arcs that are not orthogonal to the unit circle,
and horocycles are represented by Euclidean circles touching the
absolute from inside. Therefore most diagrams in the Poincar\'e model
can be drawn using Euclidean ruler and compass. However, such
constructions are not intrinsic constructions of hyperbolic
geometry. In other models, things are quite different. For example, in
the Beltrami-Klein model h-lines are Euclidean segments (chords of the
absolute), but h-circles are Euclidean ellipses.

\textit{Pseudo-elements.}  At first sight, many fundamental theorems
of Euclidean geometry fail to hold in hyperbolic geometry. For
example, in Euclidean geometry the orthocenter, the circumcenter and
the centroid of a triangle are collinear and lie on a so called
\emph{Euler line} of the triangle. Under standard definitions of
medians and altitudes, this does not hold in hyperbolic geometry
(moreover, orthocenter and circumcenter may not even exist). However,
it turns out that in the Euclidean case many notions can be defined in
equivalent ways.

For example, a median is the segment that connect a triangle vertex
with the midpoint of its opposite side, but is also a segment that
divides the triangle area in two exact halves.  In Euclidean geometry
those two notions coincide and the same term is used in both cases.
However, since in hyperbolic geometry these notions are different,
segments that connect triangle vertices with midpoints of opposite
triangle sides are usually called medians\footnote{To be more precise,
  these are h-medians as they connect vertices with h-midpoints of
  opposite sides.}, and segments that divide triangle area in halves
are called \emph{pseudomedians}~\cite{akopyan-pseudo}. Similarly,
altitudes are defined using lines perpendicular to the opposite
triangle sides. However, they could be also defined by noting that the
altitude feet $H_a$, $H_b$, and $H_c$ are three unique points on
triangle sides $BC$, $AC$, and $AB$ such that the quadrilaterals
$ABH_aH_b$, $ACH_aH_c$, and $BCH_bH_c$ are all cyclic. The second
definition gives the notion of \emph{pseudoaltitudes} (Figure
\ref{fig:pseudoaltitudes}).
 \begin{figures}[0.7]{pseudoH.tkz}
  {
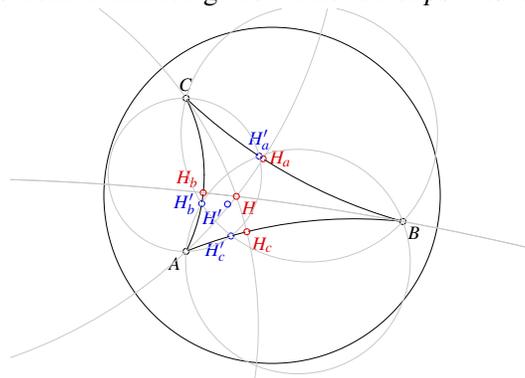
      \begin{varwidth}[t]{\linewidth}Alternative definitions of
      altitudes:\\
      orthocenter $H$ and pseudoorthocenter $H'$
    \end{varwidth}
  }
  {fig:pseudoaltitudes}
\end{figures}
The pseudoorthocenter (intersection of pseudoaltitudes), the
circumcenter, and the pseudocentroid (intersection of pseudomedians)
are always collinear and lie on a line that deserves to be called the
(pseudo) Euler line~\cite{akopyan-pseudo}. This suggests that many
Euclidean constructions that use the Euler line could be transferred
to hyperbolic geometry, if alternative definitions are used. We
advocate that hyperbolic triangle has more significant points than
Euclidean. Some significant points that are different in hyperbolic
geometry coincide in Euclidean geometry (for example, a hyperbolic
triangle has both the orthocenter and pseudoorthocenter, but they
always coincide for Euclidean triangles). Therefore, Wernick's corpus
should be extended taking into account many different
characterizations of points and notions that coincide in Euclidean,
but are different in hyperbolic geometry. This way, we may not only
consider the pseudo versions of the listed problems, but we may also
include the combinations of regular and pseudo-points (e.g. construct
a triangle given both the orthocenter and the pseudoorthocenter).

An important challenge is to determine RC-constructibility and
construction procedures for such extended set of points. For example,
while it is easy to construct the classic centroid of a hyperbolic
triangle (as an intersection of h-medians) it is challenging (if
possible at all) to construct its pseudocentroid. The same holds for
pseudoorthocenter. Since our main focus at this point is to consider
problems that can be easily RC-constructed, the pseudo objects will
not be subject of the present research.

\section{Automated triangle construction in non-Euclidean geometries}

\subsection{Primitive constructions}

Which primitive constructions can be transferred from Euclidean to
hyperbolic geometry? If we take a closer look at the list given in
Subsection~\ref{subsec:primconstE}, we see that the primitive
construction~\ref{it:w01E} cannot be carried out in the hyperbolic
case. The notion of ratio of collinear points must be substituted by
the cross-ratio of collinear points. However, in a majority of
solutions, we only use two special cases: for the fixed points $X$ and
$Y$ construct a point $Z$ that is either the midpoint of the segment
$XY$ or is symmetric to the point $X$ wrt. the point $Y$. Since these
constructions can be performed in hyperbolic geometry, they will
substitute the more general Euclidean one without much loss:

\begin{enumerate}[wide, topsep=0pt, itemsep=1pt, labelwidth=!, labelindent=0pt, label=\ref{it:w01E}\alph*.]
\item Given points $X$ and $Y$ construct the midpoint $Z$ of the
  segment $XY$;

\item Given points $X$ and $Y$ construct the point $Z$ symmetric to
  $X$ wrt.~point $Y$.
\end{enumerate}

For example, the first construction is used to construct point $M_a$
given the points $B$ and $C$, while the second one is necessary when
points $B$ and $M_a$ are given and we want to construct point $C$.

The rest of the list of primitive constructions given in
Section~\ref{subsec:primconstE} can be carried out in the hyperbolic
case as well with one exception: the primitive
construction~\ref{it:w10E} does not have a unique solution in the
hyperbolic case (there are two lines through a given point parallel to
the given line). Additionally, we can construct infinitely many lines
containing the given point that have no intersection with the given
line, i.e. the construction of the hyperparallel lines is also
possible. However, the need to construct an arbitrary line through a
point that diverges from a given line rarely arises. Usually, we need
to choose a specific line out of this set. But what choice is natural
to make? One of the characterizations of hyperparallel lines is that
they have a unique common perpendicular. Therefore, we can say that we
know how to construct a line $b\ni A$ hyperparallel to $a$ such that
the given point $A\not\in a$ is the foot of the common perpendicular
of $a$ and $b$.

Now, in the hyperbolic case, we have two primitive constructions
substituting the primitive construction~\ref{it:w10E}:
\begin{enumerate}[wide, topsep=0pt, itemsep=1pt, labelwidth=!, labelindent=0pt, label=\ref{it:w10E}\alph*.]
\item Construct the line $p$ through point $P$ that is parallel in the
  given direction to the line passing through points $X$ and
  $Y$;

\item Construct the line $p$ through point $P$ that is hyperparallel
  to the line passing through $X$ and $Y$, with
  $P$ being the foot of their common perpendicular.
\end{enumerate}

Further, we can extend the list with two new primitive constructions
that are also applicable in the Euclidean case. These were not
previously used since they were subsumed by constructions based on
stronger statements specific to the Euclidean setting (for
illustration, see Example~\ref{ex:2}). However, in absolute geometry,
it is natural to consider the image of some object under the
reflection wrt.~the given line, since every plane isometry can be
represented as a composition of a finite number of reflections. Hence,
we add the following:
\begin{enumerate}[wide, topsep=0pt, itemsep=1pt, labelwidth=!, labelindent=0pt]
\setcounter{enumi}{7}
\item Given line $m$ and point $P$, one can construct a point which is
  an image of point $P$ under the reflection wrt.~line $m$;
\item Given point $M$ and line $p$, one can construct a line which is
  an image of line $p$ under the reflection wrt.~point $M$.
\end{enumerate}

Note that the reflection wrt.~the point can be represented as a
composition of two reflections wrt.~the perpendicular lines
intersecting in that point.

\subsection{Definitions and Lemmas}\label{ssec:dl}

Definitions of the basic elements are given in terms of absolute
geometry, hence they can be transferred from the Euclidean case.
However, one should always have in mind that when saying ``line'' we
are thinking of ``h-line''.  The illustration of significant points
from Wernick's corpus in the hyperbolic setting is given in Figure~\ref{fig:characteristic_hyp}.

Although some of the lemmas from the Euclidean case can be transferred
to the hyperbolic setting (see Lemma~\ref{le:7E}), most of them needs
to be discarded or adapted.

For example, the lemmas referring to the ratio of segments, especially
the ones related to the centroid $G$ of the triangle, do not hold in
the hyperbolic case, hence the construction presented in the
Example~\ref{ex:euclid} is not possible in the hyperbolic setting.
Despite the median concurrence theorem being true, the ratios from
Lemma~\ref{le:2E} do not hold. Also, Lemma~\ref{le:1E} will hold only
in the pseudo-case.

On the other hand, in the Euclidean case the midline of the triangle
is parallel to the corresponding side of the triangle, while in the
hyperbolic case those lines are hyperparallel. Hence, the
Lemma~\ref{le:midlineE} has its hyperbolic version:

\begin{enumerate}[wide, topsep=0pt, itemsep=1pt, labelwidth=!, labelindent=0pt, label=\ref{le:midlineE}$^h$]
\item\label{le:midlineH} Lines $M_aM_b$ and $AB$ are hyperparallel and
  $M_c$ is the foot of their common perpendicular; lines $M_aM_c$ and
  $AC$ are hyperparallel and $M_b$ is the foot of their common
  perpendicular; lines $M_bM_c$ and $BC$ are hyperparallel and $M_a$
  is the foot of their common perpendicular.
\end{enumerate}
Similarly, we know that the line $AI$ is a bisector on an angle
$\angle A'AO$, where $A'$ is the foot of perpendicular from point $A$
to the midline $M_{b}M_{c}$. In the Euclidean case, points $A, A'$ and
$H$ are collinear, while in the hyperbolic one they are not. Hence,
Lemma~\ref{le:6E} cannot be used in the present form.

  \begin{figures}[0.47]{znacajneTackeHiperbolickiTrougao.tkz}
    {
      \begin{varwidth}[t]{\linewidth}
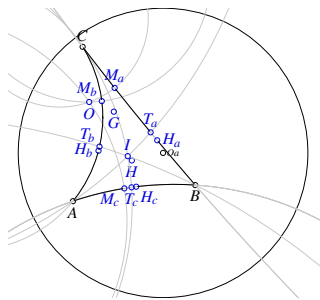
        Significant points from Wernick's corpus in hyperbolic geometry.
      \end{varwidth}
    }{fig:characteristic_hyp}
  \end{figures}

We added some more lemmas which relate to notion of reflection with
respect to point/line. All of them hold, also, in Euclidean geometry,
but we didn't need them there since we found alternative solutions to
construction problems.
\begin{enumerate}[wide, topsep=0pt, itemsep=1pt, labelwidth=!, labelindent=0pt]
\item If a vertex $A$ of triangle $ABC$ belongs to the line $p$, then
  a vertex $B$ belongs to a line which is an image of line $p$ under
  the reflection wrt.~point $M_c$;
\item Image of the vertex $B$ under the reflection wrt.~internal angle bisector $s_c$ belongs to line $AC$;
\item Image of the vertex $B$ under the reflection wrt.~internal angle
  bisector $s_c$ belongs to the circle centered at point $T_c$ passing
  through point $B$.
\end{enumerate}

\subsection{Examples}

Here, we give couple of examples illustrated in Poincar\'e disc
model. Note that all constructions can be performed using only
classical ruler and compass, since all the basic objects (h-lines,
h-circles, etc.) are either Euclidean lines or circles.

All generated solutions of problems from Wernick's corpus in the
Euclidean setting can be found on-line at
\url{http://www.matf.bg.ac.rs/~vesnap/animations/compendium_wernick.html},
and in the hyperbolic setting at
\url{http://www.matf.bg.ac.rs/~vesnap/animations_hyp/compendium_wernick_hyperbolic.html}.

\begin{example}\label{ex:1}
  Let us start with the most basic construction problem that illustrates the
  difference between Euclidean and hyperbolic geometry: Construct the
  triangle $ABC$ given three side midpoints $M_a$, $M_b$, and $M_c$.

  In the Euclidean case, we know that $AB\parallel M_aM_b$,
  $BC\parallel M_bM_c$, $AC\parallel M_aM_c$. Therefore, it is easy to
  construct lines $a:\ M_a\in a\parallel M_bM_c$,
  $b:\ M_b\in b\parallel M_aM_c$, and $c:\ M_c\in c\parallel
  M_aM_b$. Now, points $A, B,$ and $C$ are intersection points of
  lines $b$ and $c$, $a$ and $c$, $a$ and $b$, respectively (see
  \href{http://www.matf.bg.ac.rs/~vesnap/animations/construction_0341.html}{Construction
    0341}).


  As already discussed in Section~\ref{ssec:dl}, in hyperbolic geometry, the notion of parallelism can be replaced with notion of
  hyper-parallelism.  Therefore, we will use the  construction of hyperparallel lines from Lemma~\ref{le:midlineH}
  (see Figure \ref{fig:construction_0341} and \href{http://www.matf.bg.ac.rs/~vesnap/animations_hyp/construction_0341.html}{Construction 0341}).

  \vspace{0.5em}
  \begin{enumerate}[wide, topsep=0pt, itemsep=1pt, labelwidth=!,
    labelindent=0pt, leftmargin=3\parindent, label=\textbf{Step
      \arabic*.}]
  \item Construct the line $a$ that is hyperparallel to the line
    through points $M_{b}$ and $M_{c}$ with point $M_{a}$ being the
    foot of their common perpendicular;
  \item Construct the line $b$ that is hyperparallel to the line
    through points $M_{a}$ and $M_{c}$ with point $M_{b}$ being the
    foot of their common perpendicular;
  \item Construct the intersection point $C$ of the lines
    $a$ and $b$;
  \item Construct the point $B$ symmetric to $C$ wrt. point $M_{a}$;
  \item Construct the point $A$ symmetric to $C$ wrt. point $M_{b}$.
  \end{enumerate}
  \vspace{0.3em}

  \begin{figures}[0.46]{construction0_0341.tkz}
    {
      \begin{varwidth}[t]{\linewidth}
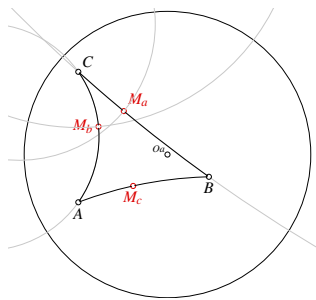
        Construction of triangle $ABC$ in hyperbolic geometry\\
        given the side midpoints $M_a$, $M_b$, and $M_c$.\end{varwidth}
    }
    {fig:construction_0341}
  \end{figures}
\end{example}

\begin{example}\label{ex:2}
  Let us consider a construction of triangle $ABC$ given vertex $A$,
  midpoint $M_a$ of side $BC$, and foot $H_b$ of the altitude from
  vertex $B$.

  This is one of the easiest problems to solve in the Euclidean case,
  requiring only four construction steps (see~\href{http://www.matf.bg.ac.rs/~vesnap/animations/construction_0044.html}{Construction 0044}).
  However, the crucial step uses the information that the inscribed
  angle subtended by a diameter is right. Unfortunately, this does not
  hold in the hyperbolic case.
  On the other hand, if we take a closer look at the problem we are trying to solve, we
  see that we actually need a way to construct a segment with
  endpoints lying on the chords of the convex angle $Opq$ (not
  necessarily the right angle) given point $M$ inside that angle as
  its midpoint. This we know how to construct: we simply need to
  construct the image $q'$ of line $q$ under the reflection wrt.~point $M$; the intersection point of $p$ and $q'$
  will give the first endpoint $P$, while the second one $Q$ will be
  symmetric to $P$ wrt.~point $M$ (see~\href{http://www.matf.bg.ac.rs/~vesnap/animations_hyp/construction_0044.html}{Construction 0044}).

  \begin{figures}[0.45]{construction0_0044.tkz}
    {
      \begin{varwidth}[t]{\linewidth}
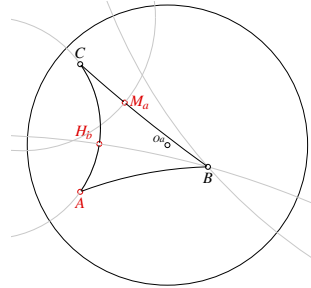
        Construction of triangle $ABC$ in hyperbolic geometry\\
        given points $A$, $M_a$, and $H_b$.
      \end{varwidth}
    }{fig:construction_0044}
  \end{figures}

  Now, the steps leading to the construction of the triangle $ABC$ in
  any geometrical setting (i.e. it holds in absolute geometry) are the
  following:

  \vspace{0.3em}
  \begin{enumerate}[wide, topsep=0pt, itemsep=1pt, labelwidth=!,
    labelindent=0pt, leftmargin=3\parindent, label=\textbf{Step
      \arabic*.}]
  \item Construct the line $b$ through the points $A$ and $H_{b}$;
  \item Construct the line $h_b$ perpendicular to the line $b$ through~$H_b$;
  \item Construct the line $s_{M_{a}}(b)$ that is image of the line
    $b$ under the reflection wrt.~point $M_a$;
  \item Construct the intersection point $B$ of the lines
    $s_{M_{a}}(b)$ and $h_{b}$;
  \item Construct the point $C$ symmetric to $B$ wrt.~point $M_{a}$.
  \end{enumerate}
\end{example}

\subsection{Results}

We considered only significantly different construction problems from
Wernick's list. Immediately, we had to discard all the problems
related to the centroid $G$ of the triangle and the ones that had to
be solved using Euler's line or Euler's circle. These problems can be
solved if we use pseudo-definitions of those objects, but in the
traditional setting, we could not find an alternate solution for
them. Solutions of all problems from Wernick's corpus we managed to
solve (including symmetric ones) in hyperbolic setting are given in
\href{http://www.matf.bg.ac.rs/~vesnap/animations_hyp/compendium_wernick_hyperbolic.html}{Hyperbolic
  corpus}. There are 31 significantly different solvable problems, 1
redundant and 11 locus dependent problems.

\section{Conclusions and further work}

We have described the first steps towards an automated system for
triangle RC-constructions in absolute and hyperbolic geometry. The
same algorithm that was previously used for Euclidean constructions
was successfully applied within our system ArgoTriCS. However,
underlying mathematical knowledge had to be substantially changed by
omitting definitions, lemmas and construction steps specific for
Euclidean geometry and by adding specifics of hyperbolic geometry.

The present experiment confirms that RC-constructions in absolute and
hyperbolic geometry are harder to make than in Euclidean geometry due
to less lemmas that can be proved about lines, circles, and
significant points of a triangle (hyperbolic triangle has more
significant points, that coincide in the Euclidean case). Also, issues
of degeneracy and existence of a solution are more complicated than in
the Euclidean case, since significant lines in a triangle may not meet
(not all pencils of lines are elliptic ones).

In our further work we shall manually investigate problems that were
not solved by our current implementation and to extend the knowledge
base (lemmas and primitive steps) so that the problems that can be
solved manually could also be solved automatically. We also
  plan to investigate potential use of our system in mathematical
  education, since studying Euclidean and hyperbolic geometry and
  their relationship can help students to acknowledge and adopt formal
  mathematics.

%

%
%
%

\bibliographystyle{eptcs}
\bibliography{hyp_constructions}

\end{document}